\documentclass[prd,twocolumn,superscriptaddress]{revtex4}
\usepackage{graphicx}

\begin{document}

\newcommand{\be}{\begin{equation}}
\newcommand{\ee}{\end{equation}}
\newcommand{\bq}{\begin{eqnarray}}
\newcommand{\eq}{\end{eqnarray}}
\newcommand{\bsq}{\begin{subequations}}
\newcommand{\esq}{\end{subequations}}
\newcommand{\bc}{\begin{center}}
\newcommand{\ec}{\end{center}}

\newcommand {\R}{{\mathcal R}}
\newcommand{\al}{\alpha}

\title{Alternatives to Quintessence Model Building}

\author{P.P. Avelino}
\email[Electronic address: ]{pedro@astro.up.pt}
\affiliation{Centro de Astrof\'{\i}sica da Universidade do Porto, R. das
Estrelas s/n, 4150-762 Porto, Portugal}
\affiliation{Departamento de F\'\i sica da
Faculdade de Ci\^encias da Universidade do
Porto, Rua do Campo Alegre 687, 4169-007, Porto, Portugal}
\author{L.M.G. Be\c ca}
\email[Electronic address: ]{lmgb@astro.up.pt}
\affiliation{Departamento de F\'\i sica da
Faculdade de Ci\^encias da Universidade do
Porto, Rua do Campo Alegre 687, 4169-007, Porto, Portugal}
\author{J.P.M. de Carvalho}
\email[Electronic address: ]{mauricio@astro.up.pt}
\affiliation{Centro de Astrof\'{\i}sica da Universidade do Porto, R. das
Estrelas s/n, 4150-762 Porto, Portugal}
\affiliation{Departamento de Matem\'atica Aplicada da
Faculdade de Ci\^encias da Universidade do
Porto, Rua do Campo Alegre 687, 4169-007, Porto, Portugal}
\author{C.J.A.P. Martins}
\email[Electronic address: ]{C.J.A.P.Martins@damtp.cam.ac.uk}
\affiliation{Centro de Astrof\'{\i}sica da Universidade do Porto, R. das
Estrelas s/n, 4150-762 Porto, Portugal}
\affiliation{Department of Applied Mathematics and Theoretical Physics,
Centre for Mathematical Sciences,\\ University of Cambridge,
Wilberforce Road, Cambridge CB3 0WA, United Kingdom}
\affiliation{Institut d'Astrophysique de Paris, 98 bis Boulevard Arago,
75014 Paris, France}
\author{P. Pinto}
\affiliation{Departamento de F\'\i sica da
Faculdade de Ci\^encias da Universidade do
Porto, Rua do Campo Alegre 687, 4169-007, Porto, Portugal}

\date{15 November 2002}

\begin{abstract}
We discuss the issue of toy model building for the dark energy component
of the universe. Specifically, we consider two generic toy models
recently proposed as alternatives to quintessence models, respectively
known as Cardassian expansion and the Chaplygin gas.
We show that the former is entirely equivalent to a
class of quintessence models. We determine the observational constraints
on the latter, coming from recent supernovae results and from 
the shape of the matter power spectrum. As expected,
these restrict the model to a behaviour that closely 
matches that of a standard cosmological constant $\Lambda$.
\end{abstract}
\preprint{DAMTP-2002-106}
\maketitle

\section{\label{intr}Introduction}

Currently available observations, especially from high-$z$ type Ia
supernovae combined with cosmic microwave background (CMB)
results \cite{Jaffe}, suggest that about one third of the critical
energy density of the
Universe is in the form of ordinary matter (including classical dark 
matter), while the remaining two thirds are in an unclustered
form which is commonly called dark energy. Among other effects,
this unknown component produces a recent accelerated expansion, 
a behaviour which standard decelerating Friedmann models are
unable to match---though see \cite{Meszaros}. The 
cosmological constant $\Lambda$ is arguably the simplest candidate 
for this dark energy, although it is well known that theoretical predictions
for its value are many orders of magnitude off from observationally
acceptable values.

Noteworthy among the many proposed alternatives are
time varying scalar fields, dubbed quintessence \cite{Caldwell,Wanga}. 
Quintessence models typically involve a scalar field function and in 
some cases more than one. However, quintessence models often suffer 
from a major problem that also afflicts the cosmological constant: 
fine-tuning. This is often referred to as the `why now?' problem: why 
is the cosmological constant (or a quintessence field) so small,
and why does it become 
dominant over the matter content of the Universe right about the 
present day? There are so-called `tracking' models where one obtains 
that quintessence energy density is reasonably independent of 
initial conditions, but on the other hand one does have to tweak some 
parameters in the scalar field potential in order to obtain the 
desired behaviour, so this can't really be claimed as an
satisfactory solution to the problem.

On the other hand, given that one has yet to see a scalar field in
action, it is clear that all such toy models are not much better justified 
than a classical cosmological constant (despite some claims to the
contrary). This is further compounded by the fact that, given some
time dependence for the scale factor and energy density, one is always able
to \emph{construct} a potential for a quintessence-type model which
reproduces them (see for example \cite{Padmanabhan}).
One is therefore reminded of Occam's razor and
can legitimately ask if observational data provide
any strong justification for them, as compared to the conceptually simpler
cosmological constant.

Here we make a contribution to this ongoing discussion, by studying
two particularly illuminating such toy models.
Cardassian expansion \cite{Freese,Freesea} has recently been
suggested as a model for an accelerating flat Universe without 
any use of a cosmological constant or vacuum energy whatsoever but 
solely depending on a purely matter driven acceleration. This has 
been accomplished by the use of a modified version of the 
Friedmann equation where an additional \emph{empirical} 
term has been added. Unfortunately, as we show here, the Cardassian 
model doesn't bring anything particularly new since, for most 
practical purposes, it reduces to a 
class of quintessence models. An example of an
alternative to quintessence is the so-called Chaplygin gas \cite{Kamenshchik},
which obeys a physically exotic equation of state (although it can be
motivated, at some level, within the context of higher dimensions 
brane theory).
In this case we will show that, as expected, current observations
restrict the model parameters to a gravitational behaviour that is
very similar to that of a cosmological constant.

\section{\label{quint}Quintessence Models}

Quintessence is a generic name given to a fluid component, other 
than ordinary pressureless matter and radiation, parameterized by an 
equation of state of the form $p_{i}=\omega_{i}\rho_{i}$ (in 
fundamental units) where $\omega_i$ is
related to a scalar field function.
Given that our chances of distinguishing between several scalar field
models yielding various different time varying
$\omega_i$ seem rather bleak \cite{Kujat,Maor},
we will for the time being take $\omega_i$ to be constant. We shall start
by reviewing some basic properties of this class of models.

\subsection{Basic Dynamical Properties}

The total energy density appearing in the 
Friedmann equation can be expanded into a sum of 
quintessential fluid components, therefore turning
\begin{equation}
\label{eq1}
{\rm{H}}^2  = \frac{8\pi G }{3}\rho  - \frac{k}{a^2}\,,
\end{equation}
where $a$ is the scale factor, ${\rm{H}} = {\dot a}/a$, the dot stands for
time derivative and $k$, the curvature term, into
\begin{equation}
\label{eq2}
{\rm{H}}^2  =  {\sum\limits_i {\Omega _i 
\frac{{\rho _i }}{{\rho _{i0} }} + 
 \left( {1 - \sum\limits_i {\Omega _i } } \right)a^{-2}} } 
\,.
\end{equation}
Here $\Omega _i  = \rho _{i0} /\rho _{c0}$ is the $i$-th 
density parameter ($\rho_{c}$ being the critical density) and the 
subscript 0 refers to present time. We have also taken $a_0=\rm{H}_0=1$.

Now, an adiabatic flow is usually assumed for each individual fluid 
component, so that $d\left( {\rho_i a^3 } \right) + 
p_id\left( {a^3 } \right) = 0$. This 
in turn implies, for each fluid
\begin{equation}
\label{eq4}
\rho _i  =   {{{\rho _{i0}}}{a^{-3(1 + \omega_i )}}
}  = \rho _{i0} \left( {1 + z} \right)^{3(1 + \omega _i 
)}\,, 
\end{equation}
where $z=1/a-1$ is the usual redshift parameter. Of course, for 
$\omega_{i}=0$ and $\omega_{i}=1/3$ we retrieve the usual relations 
for dust and radiation. Inserting this adiabatic 
condition into Eq.~(\ref{eq2}) we finally obtain
\begin{equation}
\label{eq5}
{\rm{H}}^2 ={\sum\limits_i { 
{{{\Omega_i}}{a^{-3(1 + \omega _i )}}} +  
  \left( {1 - \sum\limits_i {\Omega _{i} 
} } \right)a^{-2}} } \,.
\end{equation}
Notice also how for $\omega_{i}=-1$ a constant $\Omega_i$ term 
appears, formally acting as a $\Lambda$ cosmological constant.

A physical restriction on $\omega_i$ stems from
the square of the sound speed velocity of 
each fluid component, given by $\partial p_i /\partial\rho_i$
for constant entropy. 
The speed at which information 
is carried by the fluid must
necessarily be smaller or equal to unity, so that $\omega_i \le 1$ 
in the present case. We emphasize that this constraint is only valid for 
constant $\omega_i$ since in general $\partial p_i /\partial\rho_i$
will involve an 
additional term due to a possible dependence of $\omega_i$ on $\rho_i$. 

The dynamics of a collection of two or more different fluids can easily be 
studied if we interpret the Friedmann equation as an energy integral 
of motion of a 
one-dimensional fictitious particle moving with an $a$ coordinate. 
Consider Eq.~(\ref{eq2}) rewritten in the following form
\begin{equation}
\label{eq7}
\dot a^2-\sum\limits_i {\Omega _i}a^{ - \left({1+ 3\omega 
_i}\right)}= \left( {1 - \sum\limits_i{\Omega _i}}\right)\,,
\end{equation}
and compare it to the standard equation of motion, $E=K+V$, of the 
fictitious particle. This 
allows us to relate the curvature term to the mechanical energy 
of the system, $\dot a^2$ to its kinetic energy and the remaining 
term to the potential felt by the particle. We can
push forward this Newtonian analogy by calculating the force $-dV/da$ 
felt by the particle and see the Raychaudhuri equation emerge. 
Each fluid then contributes with a partial force of
\begin{equation}
\label{eq8}
-dV_i/da=- (1 + 3\omega _i )\Omega _i a^{ - (2 + 3\omega _i )}\,,
\end{equation}
so that fluids with $-1/3<\omega<1$ decelerate expansion while 
fluids with $\omega<-1/3$ accelerate it.

\subsection{The Cardassian Model in a Quintessence Framework}

Consider the Friedmann equation for a flat matter dominated 
Universe, \textit{i.e.}, ${\rm H}^2 = A\rho_m$, where 
$A=8\pi G /3$. Since $\rho_m \propto a^{-3}$ we 
have $\dot a^{2} =  A\rho_{m0} a^{-1}$, hence a decreasing function of the 
scale factor. Now, the easiest way of making this decreasing 
function of the scale factor a growing one is by empirically adding some 
other function that counters the $a^{-1}$ decreasing term. 
Arguably, the simplest 
function one can add is a power law of the scale factor, $a^{m}$ 
with $m>0$; let us choose $m=2-3n$. So 
now we have the following form for the time derivative of the scale factor
\begin{equation}
\label{eq11}
\dot a^{2}=A\rho_{m0}a^{-1}+Ba^{2-3n}\,,
\end{equation} 
which we can easily show to be re-writable as
\begin{equation}
\label{eq12}
{\rm{H}}^{2}=A\rho_m+B'\rho_m^{n},
\end{equation}
where $B'=B/\rho_{m0}^n$. This is precisely the
Cardassian model which consists of a 
modification to the Friedmann equation motivated by theories with extra 
dimensions.

However, since $\rho _m /\rho _{m0} = a^{-3}$ we can immediately 
rewrite Eq.~(\ref{eq12}) in the form
\begin{equation}
\label{eq9}
{\rm{H}}^2=\Omega _m a^{-3} + \Omega _\omega a^{-3n} \,,
\end{equation}
where we have defined $A =  \Omega _m /\rho _{m0} $, $B' = 
\Omega _\omega  /\rho _{m0} ^n $ and  $n=1+\omega$ (note that 
$\Omega_m+\Omega_\omega=1$). 

Notice the equivalence between equations (\ref{eq12}) and (\ref{eq9}). We can 
interpret the Cardassian empirical term in the modified Friedmann equation 
as the superposition of a quintessential fluid with 
$\omega=n-1$, to a background of dust. Note that since $m>0$ then 
$n<2/3$, implying $\omega<-1/3$.

We thus see that any Cardassian model can readily be expressed as a
quintessence model with constant $\omega$ and is therefore, for most
practical purposes, indistinguishable from it. We say `for most practical
purposes' due to the fact
that Cardassian models do not specify the behaviour of cosmological density
fluctuations on scales larger than the horizon, although it is assumed that
Newtonian gravity holds on small scales. Cardassian models are therefore
incomplete, effective toy models which describe the \textit{average universe},
which is one needs for most practical purposes. On the other hand, if one
wanted to calculate, \textit{e.g.} to calculate the CMB anisotropy on
large angular scales, one would need to go beyond this simplified procedure.
it is also worth emphasizing that one can't meaningfully
claim that one model is much better justified than the other \cite{Freese},
at least at this stage, since both are no more than toy models.

\section{The Chaplygin Gas Model}

There are various ways in which one can explain the transition from
a matter dominated Universe 
to one with an accelerated expansion using fluid components with more 
exotic state equations. Examples include brane-inspired models 
\cite{Branescan} and vacuum metamorphosis \cite{Bassett}; another 
is the generalized Chaplygin 
gas \cite{Kamenshchik} obeying an equation of state of the form
\begin{equation}
\label{eq13}
p=-\frac{A}{\rho^\alpha},
\end{equation}
where $A$ is a positive constant and $\alpha\geq 0$. The original 
form of the Chaplygin gas \cite{Kamenshchik,Bilic} where $\alpha=1$ can be 
motivated by considering a $d$-brane in a space-time of $d+2$ 
dimensions; no such motivation exists for $\alpha\neq1$. 
In this case its Nambu-Goto action can be seen as
describing a Newtonian fluid with an 
equation of state given by Eq.~(\ref{eq13}). We should note that as 
of now the Chaplygin gas is the only fluid known to admit a
super-symmetric generalization. At a toy model level, one can
trivially generalize this to a different dependence on the
density, as described in \cite{Kamenshchik,Bento}.
We emphasize that, at this simple toy-model level, this should
be seen as a generalization simply for the purpose of phenomenological
analysis.

As discussed in the above references, the Chaplygin gas can be described
in terms of a scalar field with a given potential. As
for the quintessence case, we're able to retrieve an integral of 
motion from energy-momentum conservation,
\begin{equation}
\label{integral}
d\left[ {a^{3(1 + \alpha 
)} \left({\rho^{1 + \alpha }-A}\right)}\right] = 0\,,
\end{equation}
and from it obtain the equivalent expression to Eq.~(\ref{eq4}), namely
\begin{equation}
\label{eq14}
\rho  = \rho _0 \left[ {\overline{A} +  (1 - \overline{A})
a^{-3(1 + \alpha )} } \right]^{1/1 + \alpha},
\end{equation}
where $\overline{A}=A/\rho_0^{1+\alpha}$. This quantity can be 
related to the Chaplygin gas sound speed which we can easily show 
to be $v_s^2=\alpha A /\rho^{1+\alpha}$ so that, at present, we 
have $v_{s0}^2=\alpha \overline{A}$. 

\begin{figure}
\includegraphics[width=3in,keepaspectratio]{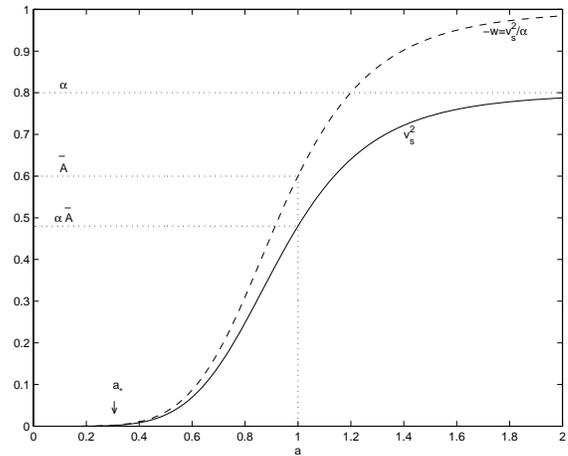}
\caption{\label{vel}
Sound speed and $\omega=p/\rho$ evolution for the case of a generalized
Chaplygin gas---we have taken $\alpha=0.8$ and $\overline{A}=0.6$,
but this behaviour is generic.
Notice the phase transition from dust ($\omega=0$) to a
cosmological constant ($\omega=-1$). The accelerating
regime begins at $a=a_*$, and the present epoch corresponds
to $a=a_0=1$.}
\end{figure}

Again, restrictions must be imposed on the values of $\alpha$ since 
we expect this velocity to be limited by the speed of light. 
In Fig.~(\ref{vel}) we find the sound speed evolution with the scale factor
for a generalized Chaplygin gas.
We note that it is bounded by $\alpha$ when $a\rightarrow \infty$;
hence $\alpha \overline{A}\leq \alpha\leq 1$ implying $\alpha
\leq 1$ and $\overline{A} \leq 1$. If we assume the generalized Chaplygin gas
to be a good approximation of the Universe for all times then these later constraints
apply. On the other hand, if we chose to make no assumptions about the future behaviour of the
Universe and treat the Chaplygin gas only as an \emph{effective} model in the sense of
effectively reproducing the Universe today without assuming anything beyond, then only 
$\alpha \overline{A} \leq 1$ applies. We will take this approach and use 
supernovae results in order to constrain the values of $\alpha$, in
the phenomenological spirit discussed above.

One of the most interesting characteristics of the Chaplygin gas 
resides in its ability to mimic the cosmological constant when 
$\overline{A}=1$ and ordinary matter when $\overline{A}=0$. We note that
the accelerating regime begins after $a_*$ given by
\begin{equation}
\label{eq15}
a_*  = \left[ {(1 - \overline{A})/2\overline{A}} \right]^{1/3(1 
+ \alpha )}\,.
\end{equation}
Thus, for $a\gg a_*$ we're able to expand the 
energy density in Eq.~(\ref{eq14}) into $\rho\approx B + C 
a^{-3(1+\alpha)}$ where $B$ and $C$ are constants. What this tells us 
is that the Chaplygin gas in the accelerating regime describes a kind 
of mixture between a cosmological constant and a type of matter known 
as stiff matter obeying the equation of state 
$p=\alpha\rho$. Note that this is again very similar to the quintessence 
results, though not quite identical:
a Chaplygin cosmology can be interpreted as an 
interpolation from a dust to 
a de Sitter Universe. This we can see most clearly in Fig.~(\ref{vel})
where we have also drawn the evolution of $\omega=p/\rho$ for a generalized
Chaplygin gas which resembles a phase transition studied in \cite{Bassett}.

\section{\label{const}Constraints from Observational Results}

We have shown the Cardassian model to be formally equivalent to a 
quintessence model if we take $\omega=n-1$. Therefore, it is 
possible to use recent constraints on $\omega$
to place limits on the Cardassian model parameter $n$. 
Several papers using supernova data have constrained the $\omega$ and 
the dark energy density, namely 
\cite{Perlmuttera,Weller,Huterer,Wangb}. Some even introduced a 
spatial (or temporal) dependency on $\omega\equiv\omega(z)$ and tried 
to ascertain the first expansion coefficients \cite{Goliath}. 
However, this work is still severely impaired by the small data set 
available today.

For a flat Universe, these analysis determined with a $95\%$ 
confidence level that $\omega<-0.45$ which includes the $\Lambda$ 
cosmological constant. On the other hand, CMB restrictions 
\cite{Melchiorri} place $\omega$ much closer to a $\Lambda$ 
scenario, in fact, with a $68\%$ confidence level $\omega$ should be 
smaller than $-0.85$. Without further observational data, no 
significantly different analysis from the work by previous authors 
can be done. Therefore, we concentrate upon the addition of a 
Chaplygin gas to a standard model.

We should also point out that a conceptually very different
possibility for a redshift-dependent equation of state
is that of an inhomogeneous universe, for example one permeated
by a network of light domain walls, which divides it into regions
with different cosmological parameters \cite{Joana}. Since these
regions tend to become more and more different as the universe 
evolves, it turns out to be quite difficult to exclude this type of
scenario, or even distinguish it from the standard one. In
particular, the presently available supernovae data still allow for
significant variations in the matter and vacuum densities \cite{Spatial1},
whereas the CMB constraints are somewhat more restrictive \cite{Spatial2}.
Note that in this type of model the equation of state changes abruptly
as the line of sight goes across a domain wall separating two such
regions. In this sense, this model is similar, for observational
purposes, to vacuum metamorphosis \cite{Bassett}.

\subsection{Type Ia Supernovae}

Following the release of the results from the Supernova Cosmology 
Project (SCP) \cite{Perlmutter} and the High-$z$ Supernova Search 
Team (H$z$ST) \cite{Riess} there has been a surging interest in the 
study of the energy content of the Universe using Type Ia Supernovae. 
Here we'll use the combined observational results from both groups,
using a procedure first described in \cite{Wang}, to produce 
constraints on the parameters of the Chaplygin gas---much simpler
estimates were done in \cite{Fabrisb}.

\begin{figure}
\includegraphics[width=3in,keepaspectratio]{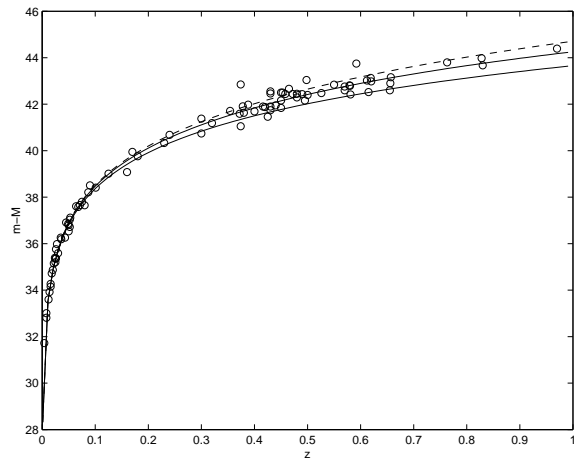}
\caption{\label{hubble}
The Hubble diagram for two fluid models of matter and Chaplygin gas.
A flat Universe with $h=0.652$ was assumed. Small circles depict our
dataset of 92 supernova measurements (error bars not displayed).
Three limiting cases are show.
The lower solid line corresponds to the $\overline A=0$ case,(matter only),
the upper solid line is the $\overline A=1$, $\Omega_m=0.3$ case (ordinary 
and dark matter plus a cosmological constant) and the
dashed line corresponds to the $\overline A=1$, $\Omega_m=\Omega_b=0.04$
(baryonic matter only, plus a cosmological constant). General 
($\overline A$,$\alpha$) models for $\Omega_m=0.3$ lie between the solid
curves while for $\Omega_m=\Omega_b=0.04$ the region extends to the dashed curve.
The matter plus cosmological constant provides a good fit so that good
Chaplygin models should 
approach it. This means that high $\overline A$ models are preferred. On the
other hand, the pure baryonic case is significantly disfavoured.}
\end{figure}

As usual, the parameter fit is based upon the luminosity distance 
$d_L$ defined through
${\mathcal F} = {\mathcal L}/4\pi d_L{}^2$ where
${\mathcal L}$ stands for the intrinsic luminosity of the 
source and ${\mathcal F}$ for the measured flux. From the Friedmann metric 
\cite{Hogg} it follows that the luminosity distance, for a flat 
geometry, as a function of redshift is given by
\begin{equation}
\label{eq17}
d_L  = d_{\rm{H}} (1 + z)\int_0^z {\frac{{dz'}}{{{\rm{E}}(z')}}}\,, 
\end{equation}
where $d_{\rm{H}}$ is the Hubble distance ($c/\rm{H}_0=1$ in 
geometrized units) and
\begin{equation}
\label{eq18}
{\rm{E}}^2(z) =\Omega _m x^3 + \Omega _{cg} \left( 
{\overline{A}  + \frac{(1-\overline{A})}{x^{-3(1+\alpha)}} } 
\right)^{1/1 + \alpha } 
\end{equation}
where $x=1+z$, for the case of a mixture of two fluids: one of ordinary matter and 
the other of generalized Chaplygin gas. The apparent magnitude $m$ of 
a supernova (a parameter more often used than the measured flux 
${\mathcal F}$ to which it is related) at a given magnitude is then 
given by
\begin{equation}
\label{eq19}
m=M+5\ \log\ d_L+25\,,
\end{equation}
$M$ being the absolute magnitude of the supernova (related to its 
intrinsic luminosity ${\mathcal L}$). Following Wang \cite{Wang}, we 
use results from both the SCP and the H$z$ST even though their 
published data sets differ in their presentation. We define the 
distance modulus to be
\begin{equation}
\label{eq20}
\mu_0=5\ \log\ d_L+25\,,
\end{equation}
as presented in the H$z$ST comprising $50$ supernovae. Comparatively, 
SCP published its measured effective rest-frame $B$-magnitude 
$m_b^{\rm{eff}}$ for $60$ supernovae which relate to the H$z$ST 
results through
\begin{equation}
\label{21}
m_b^{\rm{eff}}=M_b+\mu_0\,,
\end{equation}
where $M_b$ is the peak $B$-band absolute magnitude of a standard 
supernova. The published results of the SCP and the H$z$ST groups 
have $18$ common supernovae, $16$ of which are from the 
Cal\'an-Tololo Survey \cite{Phillips}. If we calculate $M_b$ by 
comparing results from these $18$ supernovae (using the results from 
the H$z$ST estimated by means of the MLCS method), we're able to get
\begin{equation}
\label{eq22}
M_b=m_b^{\rm{eff}}-\mu_0=-19.33\pm0.25\,.
\end{equation}
Hence, assuming the value $M_b=-19.33$ for the absolute magnitude of 
the supernovae, we can convert SCP results to distance modulus 
through Eq.~(\ref{21}). We then add $42$ of these supernovae to the 
data set from H$z$ST leaving out the $18$ already present, thus 
making a total of $92$ supernovae used.

Our dataset is displayed in Fig. \ref{hubble}, together with the
ranges which the models under discussion can take within the
diagram. Note that for each region a larger $\overline A$
corresponds to a larger value of $m-M$ at high redshift.
The same is true for large values of $\alpha$, though this effect
is subdominant relative to that of $\overline A$. This reflects the
fact that the luminosity distance is fairly insensitive to time
variations of $w$ \cite{Maor}.

\begin{figure}
\includegraphics[width=3in,keepaspectratio]{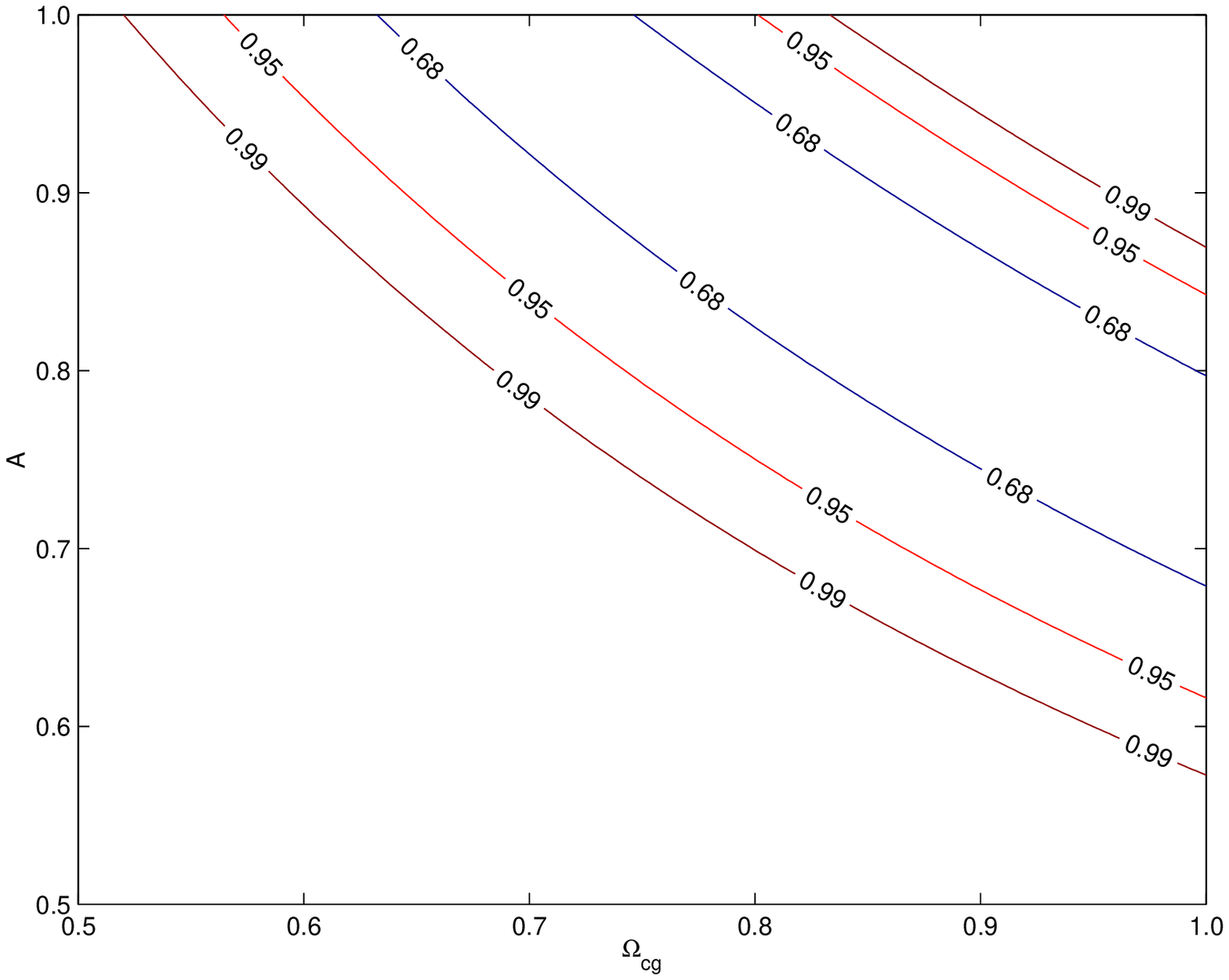}
\includegraphics[width=3in,keepaspectratio]{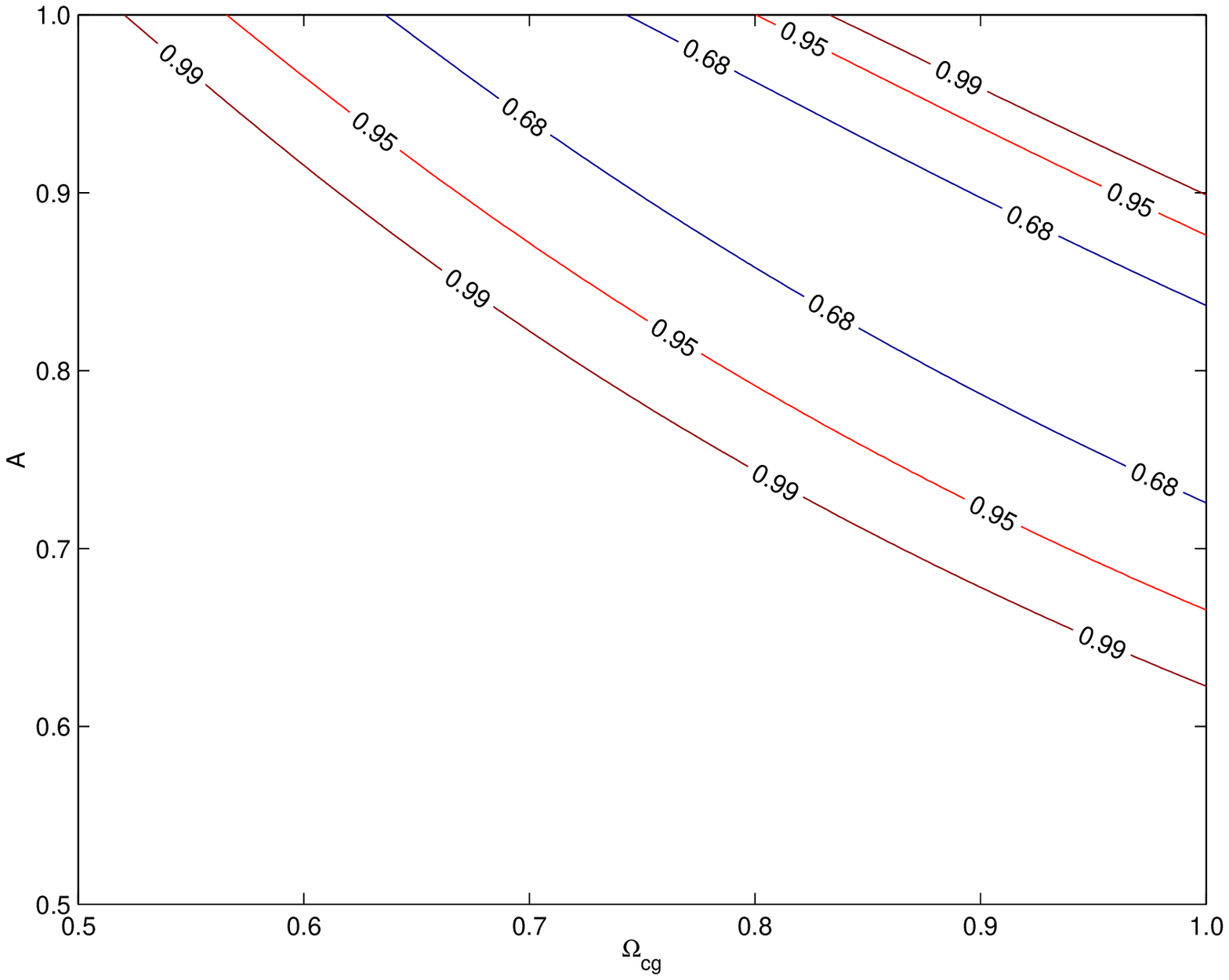}
\includegraphics[width=3in,keepaspectratio]{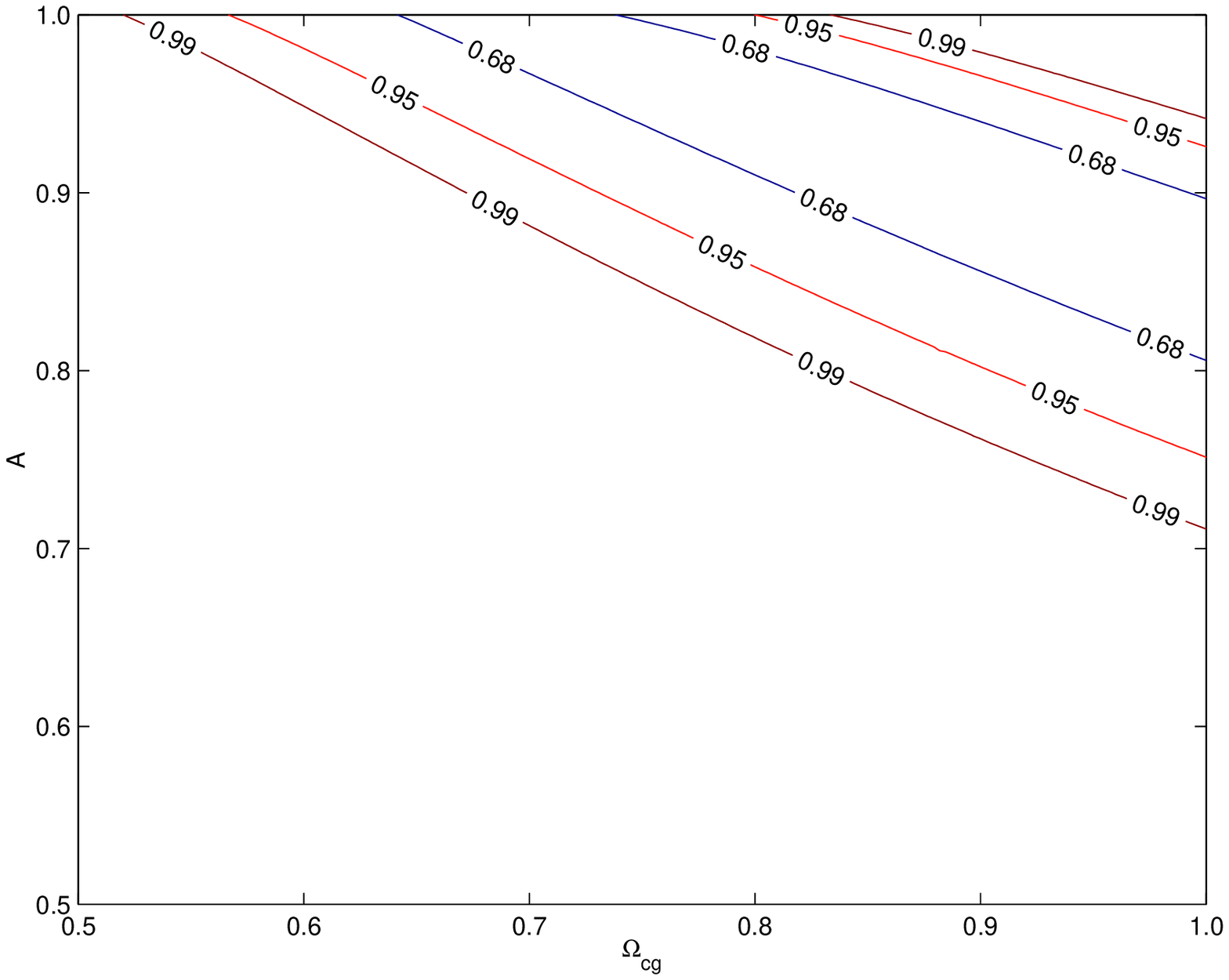}
\caption{\label{sn92_flat}
Assuming a flat Universe where only non-relativistic matter and a 
generalized Chaplygin gas are present, we show the
confidence regions resulting from a standard 
$\chi^2$ analysis fit in the $(\Omega _{cg},{\overline A})$ plane, 
using the 92 available supernovae from the combined SCP and HzST 
results for $\alpha=0.5$ (top), $\alpha=1$ (middle) and $\alpha=2$
(bottom). The Hubble 
parameter $\rm{H}_0$ has been removed by integration using standard 
procedures.}
\end{figure}

Results for the constraints imposed on the values of the Chaplygin 
gas density and its $\overline A$ parameter, for $\alpha=0.5,1,2$ 
are shown in Fig.~\ref{sn92_flat} where we have performed 
a $\chi^2$ analysis using the combined SCP and H$z$ST data set 
of the $92$ measured supernovae, assuming a flat universe filled only 
by ordinary matter and a Chaplygin gas: $\Omega_m+\Omega_{cg}=1$. 
We should point out, however,
that there are potential \textit{caveats} to the chi-squared
analysis: see for example \cite{Median}.
Here we have also integrated the Hubble parameter so as to remove its 
uncertainties from the results.

In analogy with quintessence models, where a large amount of dark 
energy with a less negative $\omega$ plays the same role as a smaller 
amount of energy with a more negative $\omega$ \cite{Perlmutter}, it 
is perceptible that the Chaplygin gas models with a higher value of 
$\alpha$ need a greater $\overline A$ to reproduce models with 
an inferior $\alpha$, even though all $\alpha$ values indicate 
roughly the same Chaplygin gas density.

The $\overline A$ parameter is restricted with a $95\%$ confidence 
level to be in the region $0.66<\overline A<1$ in the case of a 
pure Chaplygin gas form, to $0.62<\overline A<1$ for $\alpha=0.5$ and to 
$0.75<\overline A<1$ when $\alpha=2$.
Notice, however, that in the latter case the limitation 
$\alpha\overline A<1$  implies sound velocities for $\overline A>0.5$ 
greater than the speed of light. Hence the case $\alpha=2$ is in fact
completely ruled out. Similar arguments can be 
used to obtain sound velocity restrictions on the 
values of $\overline A$ and $\alpha$ (more specifically, their
product): it is often the case that the values 
given a higher probability by the $\chi^2$ analysis are physically 
unreasonable.

We should point out, however, that since one of the strongest claims of
the Chaplygin gas is that of a unified explanation for the dark matter
and dark energy, one might expect that the only components of the
universe would be the Chaplygin gas and standard baryonic matter.
In this case, since $\Omega_{b}=1-\Omega_{cg}\sim0.04$, we see that the
supernovae data strongly exclude a $\Lambda$-like behaviour, that is
the case ${\overline A}=1$.

\begin{figure}
\includegraphics[width=3in,keepaspectratio]{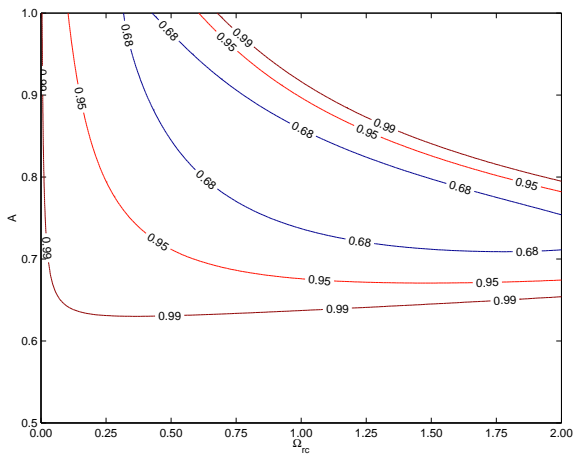}
\caption{\label{sn92_omega_b}
Assuming there is only baryonic matter and a pure Chaplygin gas, we plot the
confidence regions resulting from 
a $\chi^2$ fit in the $(\Omega_{cg},{\overline A})$ plane, using 
the 92 supernovae from the combined SCP and HzST results. A value of 
$\Omega _b = 0.04$ was assumed. Note that the 
Hubble parameter $\rm{H}_0$ has been removed by integration using 
standard procedures.}
\end{figure}

Therefore, we lift the flatness restriction 
(modifying the luminosity distance definition (\ref{eq17}) 
accordingly) and assume the energy content to be comprised of a pure 
Chaplygin gas and baryonic matter with a present day density 
of $\Omega_b=0.04$. The results of this analysis are depicted in 
Fig.~\ref{sn92_omega_b}. A large degeneracy is clearly evident.
However, the plot does seem 
to indicate that $\overline A$ should be around $\sim 0.8$ and we can 
say with a $95\%$ confidence level that $\overline A>0.69$. 
Note that for lower values of the 
Chaplygin gas density we're forced to approach the cosmological 
constant $\overline A \rightarrow 1$. Note also that for the case of a
plane geometry the acceptable values for $\overline{A}$ already exclude
a $\Lambda$ scenario. 

\begin{figure}
\includegraphics[width=3in,keepaspectratio]{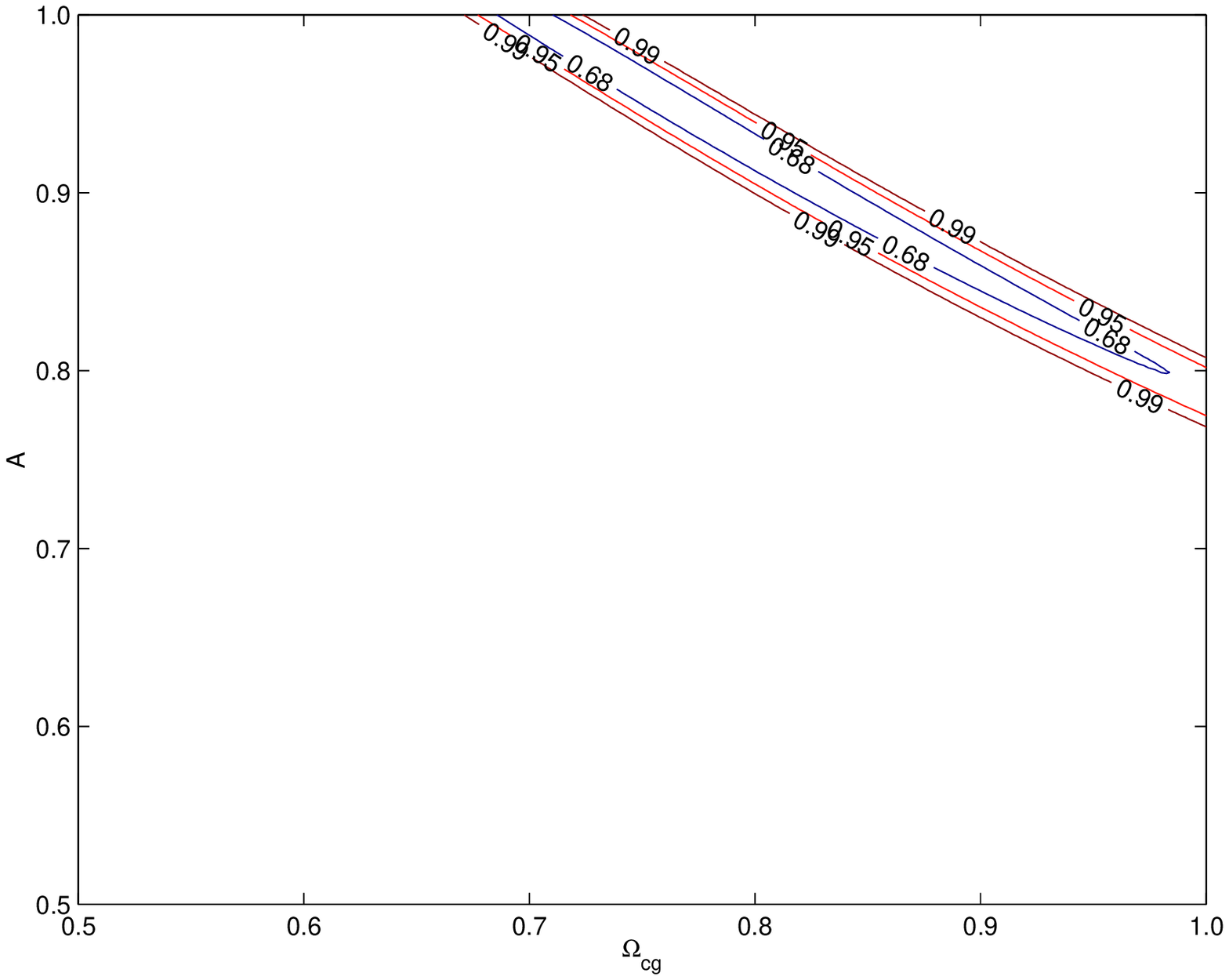}
\includegraphics[width=3in,keepaspectratio]{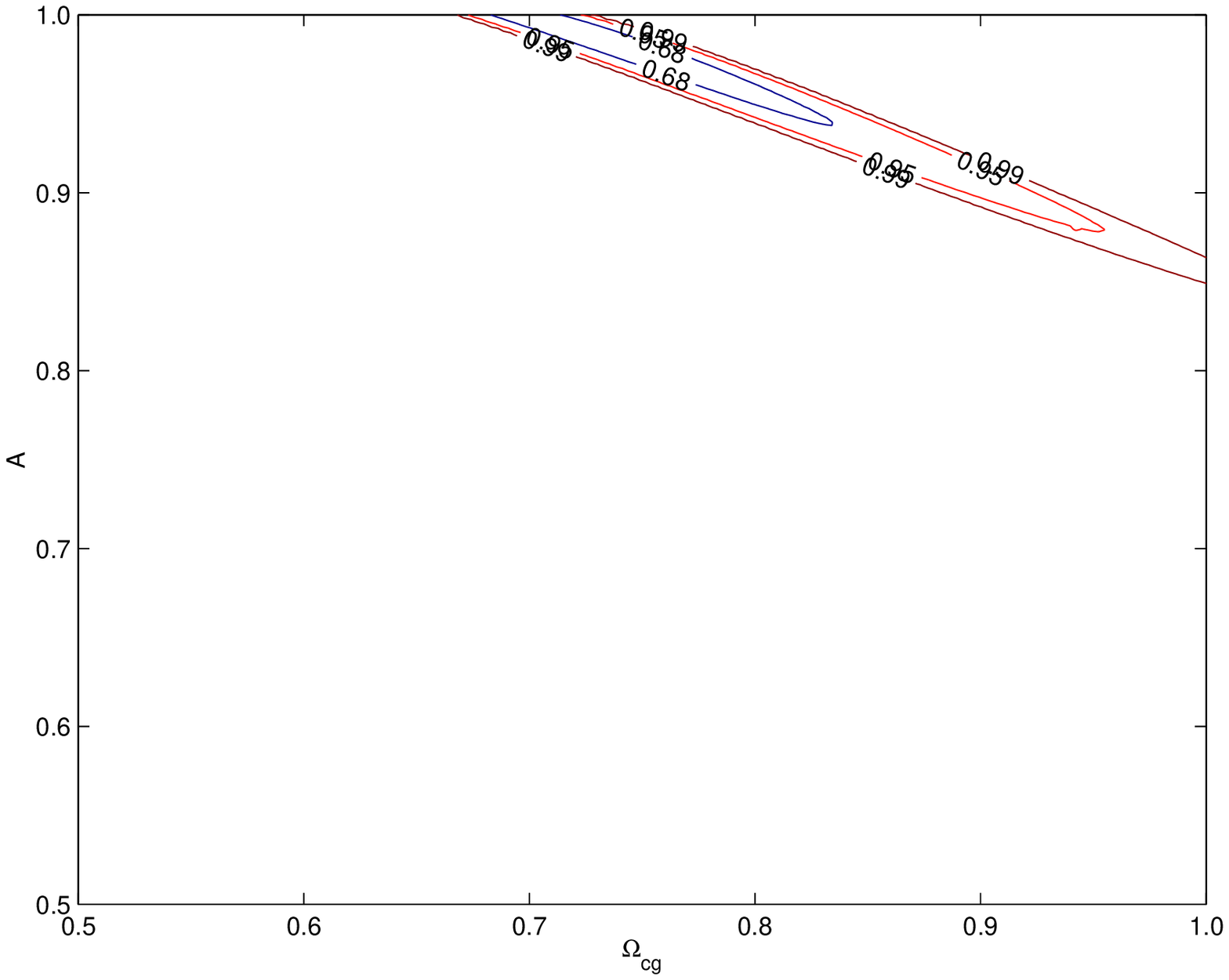}
\caption{\label{snSNAP}
A forecast of the supernovae constraints on the Chaplygin gas model,
with $\alpha=0.5$ (top) and $\alpha=1$ (bottom), for a fiducial
model $\Omega_m = 1-\Omega_{cg} = 0.3$.
Again the Hubble parameter $\rm{H}_0$ has been removed
by integration using standard procedures.}
\end{figure}

Finally we have tried to ascertain how future SNAP results will improve
our results. Assuming 
a flat Universe filled with ordinary matter and a Chaplygin gas
portrayed by $\Omega_m = 1-\Omega_{cg} = 0.3$ we repeat the $\chi^2$
analysis as detailed above, and show the corresponding
results in Fig.~\ref{snSNAP}.

Obviously the constraints are now much tighter, though it is also 
clear that there exists a fundamental degeneracy in
the $(\Omega_{cg},{\overline A})$ plane. Again we notice that for
the case $\Omega_{m}=0.3$ the model is restricted to be very similar
to a standard cosmological constant ${\overline A}\sim1$, while for
$\Omega_{m}=\Omega_{b}\sim0.04$ the $\Lambda$-like case is excluded
outright.

\subsection{The matter power spectrum}

\begin{figure}
\includegraphics[width=3in,keepaspectratio]{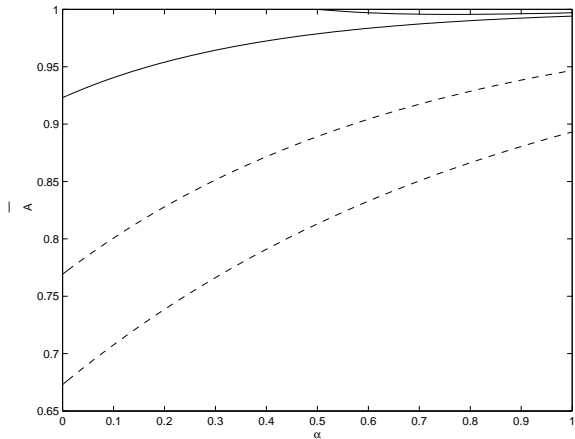}
\caption{\label{twodf}
2dF constrains on $\overline A$ as a function of $\alpha$, for
$\Omega_m=0.3$ (solid) and $\Omega_m=0.04$ (dashed). In both cases
the allowed region is enclosed by the two respective lines. A value of
$h=0.65$ was assumed for the Hubble parameter.}
\end{figure}

The length scale of the co-moving horizon at the epoch of equality
between matter and radiation is directly related to the shape of the power 
spectrum of matter perturbations. It is straightforward to show that  
this scale is proportional to the scale factor at that epoch
\begin{equation}
\label{eq24}
\frac{{{\rm{H}}_{eq}^{ - 1} }}{{a_{eq} }} \propto a_{eq}\,. 
\end{equation}
In the standard model it is a simple matter to show that 
$a_{eq}\propto(\Omega_m h^2)^{-1}$. The introduction of a generalized Chaplygin gas 
naturally implies a change in this expression which becomes
\begin{equation}  
\label{eq25}
a_{eq}  \propto \left( {\left( {\Omega _m  + \left( {1 - \Omega 
_m } \right) (1 - \overline A )^{1/1+\alpha} } \right)h^2 } \right)^{ - 
1}\,,
\end{equation}
where we have only assumed a flat universe. Note that since the Chaplygin gas behaves as
CDM except near the present time (see Fig. \ref{vel}) the shape of the power spectrum
goes unmodified by the evolution of density perturbations deep in the matter era,
except for an overall amplitude change. Indeed, the shape of the power spectrum
remains the same as for CDM except for the slight modification that Eq. (\ref{eq25})
introduces. It is common to express the wave 
number $k$ in units of $\Omega_m h^2$ or $\Gamma h$, 
where $\Gamma$ is known as the shape parameter (here we ignore the small 
correction due to the dependence on the the baryon 
density). In the case where a generalized Chaplygin gas is also present, we can show
by using Eqs~(\ref{eq24}-\ref{eq25}) that the 
shape parameter reduces to
\begin{equation}
\label{eq26}
\Gamma= \left( {\Omega _m  + \left( {1 - \Omega _m } 
\right)(1 - \overline A )^{1/1+\alpha} } \right)h\,.
\end{equation}
Recent work by \cite{Percival} using data from the 2dFGRS galaxy survey has 
constrained this parameter to be of the order of $\sim 0.2\pm0.03$, 
in agreement with preliminary SDSS results \cite{Dodelson}. Therefore,
this allows us to obtain simple bounds on $\overline A$ as a function
of $\alpha$, $h$ and $\Omega_m$. Fig. \ref{twodf} shows two such
bounds, for the `extreme' cases $\Omega_m=0.3$ and $\Omega_m=0.04$;
in both cases we have assumed a Hubble parameter $h=0.65$.
We see that $\overline{A}\geq0.65$ for all generalized 
Chaplygin models, a result indicative of the preference for $\Lambda$-type
scenarios. In the best-motivated case $\alpha=1$, the constraint is
much tighter, $\overline{A}\geq0.85$.

\section{Results and Comments}

We have studied some cosmological implications of two dark energy 
models presented as an alternative to the now standard cosmological 
constant scenario, and discussed how these toy models relate to
the more familiar quintessence paradigm.
We have shown that the Cardassian model is, for most cosmological
purposes, a standard quintessence model, whereas the Chaplygin gas
does in principle have distinguishing characteristics.

On the other hand,
by using supernova and density perturbation growth constraints, we
have shown that any Chaplygin gas type component must have a behaviour
that is very close to that of a `standard' cosmological constant
$\Lambda$, assuming that the density of `normal' (clustered) matter
is $\Omega_{m}\sim0.3$. Of course, this result is also known to apply for the
standard quintessence models \cite{Melchiorri}.
Indeed, if by an independent method we were able to determine the 
total matter content of the Universe (including dark matter) to be 
around $\sim 0.3$, then in the context of this model we would in fact
\textit{require} a 
cosmological constant so as to account for the current observational 
results (see Fig. \ref{sn92_flat}).

Conversely, the (arguably best-motivated) case where
the matter content is entirely baryonic ($\Omega_b \sim 0.04$), so that
the Chaplygin gas provides both the dark matter and the dark energy, is
the one where the differences with respect to the standard case
would be maximal. In this case, the $\Lambda$-like limit of this model
is already strongly disfavoured by observations.
It should also be noted that future observations may be able to
provide fairly tight constraints on the exponent $\alpha$, in particular
if the degeneracies discussed above are broken.

We thus conclude that the potential relevance of the Chaplygin model,
in the sense of yielding observational consequences that are
significantly different from those of the simpler cosmological constant,
is strongly dependent on the total matter content of the universe.

\begin{acknowledgments}
We are grateful to Irit Maor and Alessandro Melchiorri for
enlightening discussions of this topic.
C.M. is funded by FCT (Portugal), under grant no. FMRH/BPD/1600/2000 .
Additional support for this project came from grant CERN/FIS/43737/2001.
\end{acknowledgments}

\bibliography{qmodels}

\end{document}